\documentstyle[epsf,twocolumn]{jpsj}

\catcode`\@=11
\def\simle{\mathrel{\mathpalette\@versim<}}   
\def\simge{\mathrel{\mathpalette\@versim>}}   
\def\@versim#1#2{\lower2.5pt\vbox{\baselineskip0pt \lineskip-.5pt
   \ialign{$\m@th#1\hfil##\hfil$\crcr#2\crcr\sim\crcr}}}
\catcode`\@=12

\title
{Antiferromagnetism of the Hubbard Model\\
 on a Layered Honeycomb Lattice\\
 -- Is MgB$_2$ a Nearly-Antiferromagnetic Metal? --}

\author{Nobuo {\sc Furukawa}}

\def\runtitle{ A Hubbard Model on a Layered Honeycomb Lattice}
\def\runauthor{Nobuo {\sc Furukawa}}

\inst
{
Department of Physics, Aoyama Gakuin University, 
Setagaya, Tokyo 157-8572
}

\recdate
{ 
March 12, 2001
}

\abst
{
As a model for the B 2$p_z$ bands of MgB$_2$, 
a half-filled Hubbard model on a honeycomb lattice is studied,
and a possibility 
of an antiferromagnetism (AF) at the ground state is examined by Monte Carlo 
calculations as well as a random phase approximation method.
On a single-layer honeycomb lattice,
a paramagnetic to AF phase transition occurs
at a finite value of the repulsion energy $U_{\rm c}/t \simeq 3.6$,
due to the loss of the density of states at the fermi level.
On a layered honeycomb lattice, however,
inter-layer hopping term creates a perfect nesting of the fermi surface,
and leads the AF with $U_{\rm c}=0$.
From these facts, we discuss that MgB$_2$ is possibly a 
nearly-AF metal, where the AF of
the  2$p_z$ 
bands is destroyed by carrier doping.
}

\kword
{
Hubbard model, honeycomb lattice,
 MgB$_2$, superconductivity, antiferromagnetism, nesting
}

\begin{document}
\sloppy
\maketitle

\section{Introduction}
The recent discovery of MgB$_2$,\cite{Nagamatsu01} 
an intermetallic superconductor 
with high superconducting transition 
temperature ($T_{\rm c}$), initiated a large number of activities both in   
experimental and theoretical researches.
Several experiments, including NMR,\cite{Kotegawa01x}
specific heat,\cite{Walti01x} as well as
photoemission spectrum\cite{Takahashi01x} measurements
suggest that this material shows BCS-like $s$-wave superconductivity.
Band 
calculations\cite{Kortus01x,Belashchenko01x,Satta01x,An01x,Kong01x} 
indicate a strong interaction between electrons and 
high-frequency phonon modes exists,  which has been discussed as a 
possible origin of high $T_{\rm c}$ superconductivity in this material.
It has been also argued that roles of strong Coulomb interactions may 
directly or indirectly  give rise to high 
$T_{\rm c}$.\cite{Hirsch01x,Imada01x,Voelker01x}

The crystal of MgB$_2$ consists of alternative stackings of  
honeycomb-lattice B$_2$ layers and triangular-lattice Mg layers,
as is in AlB$_2$.
The band calculations show that Mg are almost fully ionized, and the
bands at the fermi level have mostly B 2$p$ characters.
There exist four fermi surfaces, two of them in the 2$p_z$ bands
and the rest in the 2$p_{x,y}$ bands.
The 2$p_{x,y}$ bands are derived from the intra-layer
 $\sigma$ bonding orbitals, and have a quasi two-dimensional hole 
character.
The  2$p_z$ bands are derived from
the intra-layer $\pi$ bonding orbitals which also have
inter-layer couplings between adjacent atomic orbitals 
in the $c$-axis direction.

\begin{figure}[htb]
\hfil\epsfxsize=4cm\epsfbox{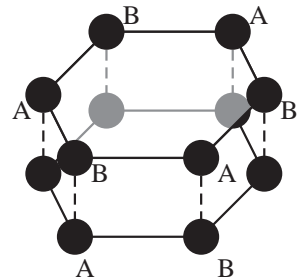}\hfil
\caption{Boron sites in the layered honeycomb
 lattice structure of MgB$_2$.
Tight-binding fits show that a N.N.
hopping model between boron sites reproduces  the  2$p_z$ bands. 
Characters A and B denote sublattice indices.}
\label{FigHexa}
\end{figure}

The tight-binding fits to the band structure
 calculations\cite{Kortus01x,Voelker01x} show that
the 2$p_z$ bands are well reproduced by
a nearest neighbor (N.N.) hopping model on a layered honeycomb lattice
of B sites in the  formulae
\begin{eqnarray}
  \varepsilon_{\pm}({\bf k}) &=&
   \pm 2 t_{xy} \gamma_{xy}({\bf k})
 - 2 t_z \cos(k_z) ,
  \label{dispersion}\\
 \gamma_{xy}({\bf k}) &=&
   \sqrt{
    \frac34 + \frac12 \cos(k_x)+ \cos(\frac12k_x)\cos(\frac{\sqrt3}2 k_y)
   },
   \label{stfactor}
\end{eqnarray}
where $t_{xy}$ ($t_z$) denotes intra-layer (inter-layer) N.N. electron
hopping energy, while $\gamma_{xy}$ is the structure factor of the
intra-layer electron hopping to the N.N. sites on a honeycomb lattice.
A lattice structure of the tight-binding model is
illustrated in Fig.~\ref{FigHexa}.

In this Letter, we investigate an aspect of the  2$p_z$ bands
in MgB$_2$
which comes from its characteristic band structure.
We consider a Hubbard model on a layered honeycomb lattice.
Using a random phase approximation (RPA) as well as
quantum Monte Carlo (MC) methods,
antiferromagnetism (AF) at the ground state 
is examined.
We discuss that inter-layer electron hoppings 
create a perfect fermi-surface nesting in the 2$p_z$ bands.
With infinitesimal values of the Coulomb repulsion,
the system is unstable against an AF order at half filling.

First, let us consider a half-filled Hubbard model on a general 
bipartite lattice with N.N. hoppings.
This model has a particle-hole symmetry, and
as a result, a non-interacting
electron dispersion has a relation
\begin{equation}
\epsilon(k+Q) = -\epsilon(k) ,
  \label{PHenergy}
\end{equation}
where $Q$ is the AF wavevector defined
by the bipartite structure of the lattice,
{\em e.g.} $Q=(\pi,\pi)$ for a square lattice.
To be more precise,
the relation (\ref{PHenergy}) holds when a electron on
a A-sublattice site hops only to sites on B-sublattice, and vice versa.
A particle-hole symmetry gives the chemical potential $\mu=0$ at half filling.
On the fermi surface, the relation
 $\varepsilon(k)=\varepsilon(k+Q)=0$ means that
there exists a perfect nesting with
the nesting vector $Q$.

Upon the presence of the particle-hole symmetry,
non-interacting magnetic susceptibility at $Q$ is given by
\begin{eqnarray}
 &&\chi_0(Q) \equiv -\frac1N \sum_k
    \frac{ f(\epsilon(k)) - f(\epsilon(k+Q))} 
{\epsilon(k) - \epsilon(k+Q) }
   \nonumber \\
&&\quad=
-\int D(\varepsilon) 
{\rm d}\varepsilon 
\frac{ f(\varepsilon) -\frac12}{\varepsilon}
=
 -\int D(\varepsilon) 
{\rm d}\varepsilon 
\frac{ f(\varepsilon)}{\varepsilon},
\label{chi0Q}
\end{eqnarray}
where $f(\varepsilon)$ is the fermi distribution function.
In the derivation, we used that $D(\varepsilon)/\varepsilon$ is an
odd function of $\varepsilon$ so that
 its principal value integration vanishes.
Generally, when  density of states (DOS) at the fermi level is finite,
the susceptibility $\chi_0(Q)$ shows a logarithmic 
divergences.
On a square lattice,  van Hove singularity
creates a divergence of the DOS, $ D(\varepsilon)\propto 
\log(\varepsilon)$,
and hence we have a log-square singularity in $\chi_0(Q)$.

Using an RPA,
the criterion for the AF
 instability is given by $1-U\chi_0(Q) = 0$, where $U$ is the
on-site repulsion energy.
The divergence of $\chi_0(Q)$ implies
that an AF  spin density wave (SDW) occurs at infinitesimal 
values of $U$. In other words, the critical value for the SDW
phase transition is  $U_{\rm c}=0$.
The SDW state is smoothly connected to the strong coupling limit
$U\to\infty$, where the system is well described by
an AF Heisenberg model on a bipartite lattice.

Let us now consider a half-filled Hubbard model 
on a single-layer honeycomb lattice 
with N.N. hoppings.
The energy dispersion is given by 
$\epsilon_{\pm} (k) = \pm 2t \gamma_{xy}(k)$,
where $t$ is the hopping energy.
The energy dispersion has a particle-hole symmetry
due to the bipartite nature of the lattice.
What is unique in this lattice is that
the DOS decreases to zero at the fermi level,
\begin{equation}
  D(\varepsilon) \propto |\varepsilon| .
  \label{DOShone}
\end{equation}
This may be understood from the fact that,
at $\mu\to0$, the fermi surface shrinks to points at $K$-points
$(\pm 2\pi/\sqrt{3}, \pi)$.
Using eq.~(\ref{DOShone}) in eq.~(\ref{chi0Q}), we see that the
singularity in $\chi_0(Q)$ is suppressed. Then, RPA gives that
the AF instability occurs at a non-zero value of $U_{\rm c}= 1/\chi_0(Q)$.
The detailed calculation gives $U_{\rm c}/t=2.23$.\cite{Sorella92}

In order to examine this result with respect to
the reliability of the RPA treatment, we perform a quantum Monte Carlo
calculation of the Hubbard model on a honeycomb lattice.
We apply a projection algorithm with 
optimized initial states.\cite{Furukawa91zz}
The ground state is obtained by
 $ |\Psi_{\rm G}\rangle \simeq \exp(-\tau {\cal H}) |\Psi_{\rm 
ini}\rangle$
at large enough values of $\tau$.
Here, ${\cal H}$ is the Hamiltonian of the Hubbard model
 with electron hopping energy $t$ and a repulsive interaction $U$.
For the initial state $|\Psi_{\rm ini}\rangle$, we use
a Hartree-Fock solution for the identical lattice
with an interaction $U_{\rm ini}$.
Note that $U_{\rm ini}$ ($\ne U$) can be chosen arbitrarily,
and an appropriate value for $U_{\rm ini}$ accelerates the
convergence to the ground state.

Monte Carlo runs are performed
for various system sizes with linear dimension
$L=6$, $8$, $10$, $12$, $14$ and $16$.
We typically run 10,000 Monte Carlo steps after
1000 relaxation steps.
AF structure factor is calculated by
\begin{equation}
 S(Q) = \frac1N \sum_{i,j} S_i \cdot S_j (-1)^{r_i - r_j}.
\end{equation}
Here $N$ is the number of sites.
Staggered magnetization $M_{\rm s}$ is obtained by 
\begin{equation}
  M_{\rm s}{}^2 = \lim_{N\to\infty} S(Q)/N.
\end{equation}
Therefore, existence of the AF order 
can be
 judged by the extrapolation of $S(Q)/N$ to the thermodynamic 
limit.

\begin{figure}[htb]
\hfil\epsfxsize=8cm\epsfbox{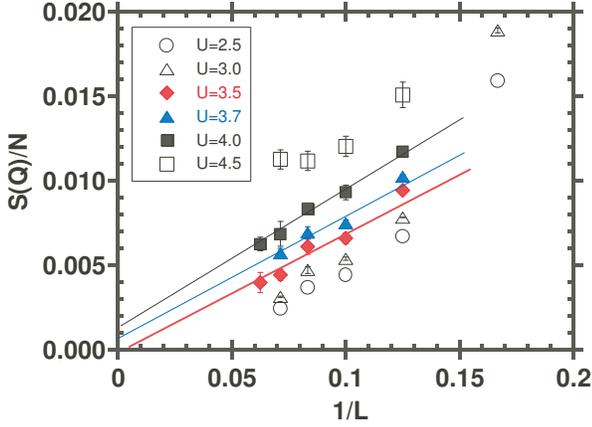}\hfil
\caption{System size dependence of the AF structure factor
for a Hubbard model on a single-layer honeycomb lattice.
Lines in the figure are least-squares fits for
$U/t=3.5$, $3.7$ and $4.0$ from the bottom to the top.}
\label{FigLinv}
\end{figure}

System size dependence of $S(Q)/N$ 
for various values of $U$ is depicted in Fig.~\ref{FigLinv}.
We extrapolate the data by $1/L$.
Here, finite size effects
due to linear spin-wave type excitations  in the AF ordered 
states are assumed.
From the data we see that the extrapolation gives
$M_{\rm s}>0$ at $U/t \ge 3.7$.
Meanwhile, at $U/t \le 3.5$ the $1/L$ extrapolation gives
negative values for $M_{\rm s}{}^2$,
indicating that  $1/L$-dependence derived from the linear spin wave 
theory
does not hold in this case.
From this fact, 
the critical value for the
onset of AF order is estimated to be $U_{\rm c}/t\simeq 3.6$.\cite{Comment1}
Thus, quantum fluctuations do not extinguish RPA instability.
They only renormalize the value of $U_{\rm c}/t$ by a factor of about 1.6.
As long as the existence of the instability is concerned,
RPA is considered to be reliable.

Next, we introduce inter-layer hoppings.
For a while, we assume a half-filled situation for
the 2$p_z$ bands, of which energy is described by 
eq.~(\ref{dispersion}).
In actual MgB$_2$, however, existence of hole fermi surfaces in
the 2$p_{x,y}$ bands leads a deviation from half-filling for 
the 2$p_z$ bands. The doping effect will be discussed later.

As illustrated in Fig.~\ref{FigHexa}, the lattice structure
is bipartite. We  take intra-layer and 
inter-layer N.N. hoppings only.
An inter-sublattice hopping on a bipartite lattice creates
a particle-hole symmetry.
Indeed, a non-interacting electron dispersion in eq.~(\ref{dispersion})
shows
\begin{equation}
  \varepsilon_{+}(k_x,k_y,k_z) = - \varepsilon_{-}(k_x,k_y,k_z+\pi).
  \label{PHsymmetry}
\end{equation}
Note that,
since a unit cell contains two boron atoms and the
 first Brillouin zone is already halved by the 
crystallographic reason,
the AF wavevector in this lattice is $Q=(0,0,\pi)$.
As an inter-layer hopping $t_z$ is increased from zero,
the energy dispersion along the $k_z$ axis makes
the fermi surface expand from the lines in the Brillouin zone
connecting K and H points,
and creates fermi surfaces with finite surface volumes
around $k_z=0$ and $k_z =\pi$.
Therefore, at $t_z \ne0$,
the DOS at the fermi level becomes non-zero.
Then, from eq.~(\ref{chi0Q}), 
$\chi_0(Q)$  diverges logarithmically,
and the RPA instability occurs at infinitesimal values of $U$.

\begin{figure}[htb]
\hfil\epsfxsize=6cm\epsfbox{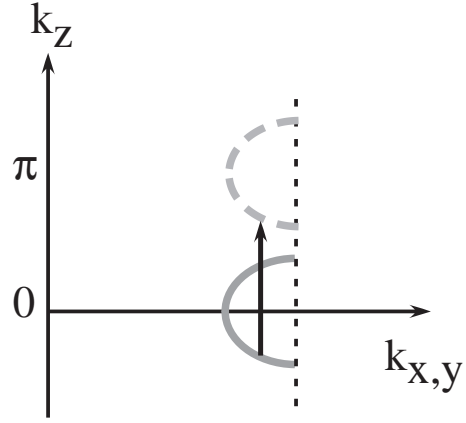}\hfil
\caption{Schematic illustration of the fermi surfaces
and the nesting vector on a layered honeycomb lattice
with finite inter-layer hopping terms.
 The hole (electron) fermi surface around $k_z=0$
($k_z=\pi$) is shown by  solid (dashed) gray curves.
The nesting vector $Q=(0,0,\pi)$ is given by an arrow.
The dashed line represents the Brillouin zone boundary.
}
\label{FigFS}
\end{figure}

This behavior can be understood as a creation of a fermi surface nesting
by increasing the inter-layer electron hoppings.
For any $k$-points on the fermi surface, the relation
$ \varepsilon_{+}(k_x,k_y,k_z) = - \varepsilon_{-}(k_x,k_y,k_z+\pi)=0$
holds, which means the perfect interband nesting with $Q=(0,0,\pi)$.
The situation is schematically illustrated in Fig.~\ref{FigFS}.
Increase of $t_z$ makes the surface volumes of the
perfectly-nested fermi surfaces increases.

Enhancement of the nesting behavior by the inter-layer hoppings
is in a great contrast with usual quasi low-dimensional systems
where three dimensionality induces curvatures of the
fermi surface and hence decreases nesting properties.
This unique feature comes from the bipartite geometry of
a honeycomb lattice.

Within this band structure,
$U_{\rm c} \ne 0$ holds strictly only at $t_z=0$,
while  we suddenly have $U_{\rm c}=0$ at $t_z\ne0$.
$U_{\rm c}$ behaves singularly as a function of $t_z$.
In other words, for a small value of $U$,
SDW order appears as soon as inter-layer hopping is allowed.
In practice, however, there exist particle-hole symmetry breaking terms,
{\em e.g.} second neighbor hoppings as well as a chemical potential
shift due to carrier doping.
Then, there exists a competition between the inter-layer hopping term
which create the fermi surface nesting and the symmetry breaking terms 
which
destroys the nesting. As a result, the singular behavior
in the $t_z$ dependence of $U_{\rm c}$ will be smeared out
by the particle-hole symmetry breaking terms, and $U_{\rm c}$
is speculated to remain non-zero. Nevertheless,
as long as  the particle-hole symmetry breaking terms are small,
we expect that AF instability occurs at relatively small values of $U$.

If the 2$p_z$ bands of MgB$_2$ are half filled, the strong nesting
behavior of the fermi surfaces creates an AF-SDW state
even if the Coulomb repulsion is not so large.
Therefore, it is possible that MgB$_2$ 
is a nearly-AF metal.
Self-doped carriers from the 2$p_{x,y}$ bands to the
2$p_z$ bands might have destroyed the AF order
and make MgB$_2$  nonmagnetic,
being similar with the case for high-$T_{\rm c}$ cuprates in the doped 
region.
At this point, however, it is not clear 
whether such AF is correlated to the mechanism of the
superconductivity in MgB$_2$, {\em e.g.} whether remnant 
AF fluctuations assist the phonon mechanism to enhance the
attractive interaction for the Cooper pairing.
Further detailed studies are required to address this question.

In this sense, lattice structure of MgB$_2$ is special because
the layer stacking conserves the bipartite nature.
As a counter example, we make a comparison with graphite, where
 stacking of the honeycomb layers 
are not similar to those for MgB$_2$. A carbon atom on
an adjacent layer lies on top of the center of a hexagon,
which is not a bipartite structure. Then, the
system is stable against AF instabilities in the small $U$ region.
This explains 
the fact that graphite does not show AF.
Coulomb interactions are expected to be not
strong enough to induce a magnetically ordered state.

In order to justify the idea that MgB$_2$ is a nearly-AF metal,
we propose some crucial tests.
From the band calculation,
more precise estimate for the band dispersion,
either by the tight-binding fits, or by directly
calculating the fermi surface, will give
how strong the nesting behavior is in this material.
Coulomb repulsion energy should also be estimated
to clarify whether it is strong enough to cause AF states
if the 2$p_z$ bands are half filled.
Experimentally, measurement of remnant AF spin fluctuations
in MgB$_2$ is important.
A search for AF states in various diboride compounds
with half-filled 2$p_z$ bands is also interesting.

To summarize,
ground state magnetism
of the  Hubbard model on a layered honeycomb lattice at half-filling
is studied as a model for the B 2$p_z$ bands of MgB$_2$.
RPA as well as MC calculations are performed
to study AF-SDW singularities.
On a single-layer honeycomb lattice,
AF instability occurs
at $U_{\rm c}/t \simeq 3.6$.
On a layered honeycomb lattice,
inter-layer hopping term induces the fermi-surface nesting,
contrary to the usual cases where three dimensionality
destroys nesting behaviors.
Within RPA scheme, the perfect nesting
leads AF to occur at $U_{\rm c}=0$.
We discuss that
MgB$_2$ might  be a nearly-AF metal.

The computations have been performed on
the massive-parallel personal computer systems  of AOYAMA+ project
(http://www.phys.aoyama.ac.jp/\~{}aoyama+).
This work is supported by a Mombu-Kagakusho Grant-in-Aid for Scientific 
Research.


\begin{thebibliography}{10}

\bibitem{Nagamatsu01}
J. Nagamatsu, N. Nakagawa, T. Muranaka, 
Y. Zenitani 
and J. Akimitsu: Nature {\bf 410} (2001) 63.

\bibitem{Kotegawa01x}
H. Kotegawa, K. Ishida, Y. Kitaoka,
T. Muranaka, J. Akimitsu: cond-mat/0102334.

\bibitem{Walti01x}
Ch. W{\"a}lti {\em et al.}:
cond-mat/0102522.



\bibitem{Takahashi01x}
T. Takahashi, T. Sato, S. Souma, T. Muranaka and J. Akimitsu:
cond-mat/0103079.


\bibitem{Kortus01x}
J. Kortus, I.I. Mazin, L.D. Belashchenko, 
V.P. Antropov and L.L. Boyer: cond-mat/0101446.

\bibitem{Belashchenko01x}
K.D. Belashchenko, M. van Schilfgaarde and V.P. Antropov:
cond-mat/0102290.

\bibitem{Satta01x}
G.Satta, G.Profeta, F.Bernardini, A.Continenza and S.Massidda:
cond-mat/0102358.

\bibitem{An01x}
 J. M. An, W. E. Pickett: cond-mat/cond-mat/0102391.

\bibitem{Kong01x}
Y. Kong, O.V. Dolgov, O. Jepsen and O.K. Andersen:
cond-mat/0102499.


\bibitem{Hirsch01x}
J.E. Hirsch: cond-mat/0102115.

\bibitem{Imada01x}
M. Imada: cond-mat/0103006.

\bibitem{Voelker01x}
K. Voelker, V.I. Anisimov and T.M. Rice: cond-mat/0103082.


\bibitem{Sorella92}
S. Sorella and E. Tosatti:
Europhys. Lett. {\bf 19} (1992) 699.

\bibitem{Furukawa91zz}
N. Furukawa and M. Imada:
 J. Phys. Soc. Jpn. {\bf 60} (1991) 3669.

\bibitem{Comment1}
This number is slightly smaller than an estimated value  by
Sorella and Tosatti in Ref.~\citen{Sorella92},  $U_{\rm c}/t \simeq 4.5$,
which was derived from charge correlation behaviors
under SDW gaps on  small clusters $L \le 8$ without
a system size extrapolation.

\end{thebibliography}

\end{document}